\documentclass[ reprint, superscriptaddress, amsmath, amssymb, aps, pra, twocolumn ]{revtex4-1}

\usepackage{hyperref}
\usepackage{xurl}
\hypersetup{breaklinks=true}

\usepackage{graphicx}
\usepackage{dcolumn}
\usepackage{bm}

\usepackage{mleftright}
\usepackage{color}
\usepackage{natbib}

\usepackage{amsmath}

\newcommand*{\defeq}{\stackrel{\text{def}}{\scalebox{1.5}[1]{=}}}

\usepackage{dsfont}

\usepackage{tikz}
\usetikzlibrary{backgrounds, fit, decorations.pathreplacing, calc}

\begin{document}

\title{Noise-tailored Constructions for Spin Wigner Function Kernels}

\author{Michael Hanks}
\email{m.hanks@imperial.ac.uk}
\affiliation{QOLS, Blackett Laboratory, Imperial College London, London SW7 2AZ, United Kingdom}

\author{Soovin Lee}
\affiliation{QOLS, Blackett Laboratory, Imperial College London, London SW7 2AZ, United Kingdom}

\author{M.S. Kim}
\affiliation{QOLS, Blackett Laboratory, Imperial College London, London SW7 2AZ, United Kingdom}

\date{\today}

\begin{abstract}
    The effective use of noisy intermediate-scale quantum devices requires error mitigation to improve the accuracy of sampled measurement distributions.
    The more accurately the effects of noise on these distributions can be modeled, the more closely error mitigation will be able to approach theoretical bounds.
    The characterisation of noisy quantum channels and the inference of their effects on general observables are challenging problems,
    but in many cases a change in representation can greatly simplify the analysis.
    Here, we investigate spin Wigner functions for multi-qudit systems.
    We generalise previous kernel constructions, capturing the effects of several probabilistic unitary noise models in few parameters.
\end{abstract}

\maketitle

\section{Introduction}
\label{sec:introduction}

Quantum technologies promise increased capabilities in a number of areas, such as
metrology \cite{taylor_quantum_2016},
cryptography \cite{shor_algorithms_1994,renner_debate_2023},
physical simulations \cite{hagan_composite_2023},
and computational fluid dynamics \cite{succi_quantum_2023}.
However, to see an advantage in many of these applications requires that the quantum device operate fault tolerantly and at a large scale
\cite{hoefler_disentangling_2023}, a challenge that is unlikely to be met with current architectures within the next five to ten years.
Recently, noisy intermediate-scale quantum (NISQ) devices \cite{preskill_quantum_2018} have received significant attention, with much effort expended to show near-term quantum advantage.
Several works \cite{arute_quantum_2019,kim_evidence_2023}
have claimed evidence of such advantage for certain highly particular problems,
but the utility of these devices remains uncertain
\cite{pan_simulating_2021,tindall_efficient_2023}.

Error mitigation methods
\cite{cai_quantum_2023}
have been developed to increase the accuracy of sampled quantities on NISQ quantum devices constrained in qubit number and circuit depth.
The characterisation of noise and its effect on a quantum state is however itself a difficult problem \cite{kaufmann_characterization_2023}.
As a result, we often find either that the noise is symmetrized (and increased) via twirling, or that error mitigation techniques such as probabilistic error cancellation \cite{van_den_berg_probabilistic_2023} or zero-noise extrapolation, which may in general involve some subtlety \cite{majumdar_best_2023}, are justified empirically rather than from first principles at the device level.

Common features of many noise channels are high levels of symmetry and the dominance of low energy components, and these properties recommend a signals-processing perspective of filters and window functions.
In the case of the harmonic oscillator, we observe that noise processes are often fruitfully decomposed
\cite{glauber_coherent_1963,kim_quantum_1996}
as either displacements, rotations, or steps in photon number, and we would like to explore similar methods of decomposition for multi-qubit systems.
Spin Wigner functions \cite{brif_phase-space_1999,tilma_wigner_2016} have been developed as finite-dimensional analogues of the well-known quasi-probability distribution for the harmonic oscillator, and have previously been proposed as a visualisation tool and for verifying the preparation of quantum states \cite{rundle_simple_2017,rundle_quantum_2017}.
Moreover, the reference frames of more discrete variants of phase-space quasi-probability distributions \cite{park_efficient_2023} have been exploited for Clifford circuit simulation.
We find in this paper that spin Wigner functions, with their significant freedom in parameterisation, can also provide a convenient representation for the treatment of probabilistic unitary noise.

The paper is laid out as follows:
Section~\ref{sec:spin_wigner_functions_and_spherical_harmonics} reviews spin Wigner functions and higher-dimensional spherical harmonics.
Section~\ref{sec:parameterisation_via_harmonics_and_unitary_designs} explores the requirements of general spin Wigner function kernels, laying out a general form in terms of the spherical harmonics and discussing several example constructions.
Section~\ref{sec:example_noise_models} turns these constructions toward the key question of noise, discussing how the form of the kernel is altered when parameterising operations occur only over a restricted subgroup or subalgebra.
Section~\ref{sec:error_mitigation_via_rescaled_expectation_values} discusses the potential for insights in error mitigation, rescaling expectation values according to the form of the noise distribution.
Finally, we conclude in Section~\ref{sec:discussion} by discussing the implications of our results, their scope, and possible next steps.

\section{Spin Wigner Functions and Spherical Harmonics}
\label{sec:spin_wigner_functions_and_spherical_harmonics}

\subsection{Spin Wigner Functions}

The spin Wigner function \cite{brif_phase-space_1999,tilma_wigner_2016},
$${
	W_{\hat{\rho}}
	\left( \xi \right)
	=
	\text{Tr}
	\left[
		\hat{\rho}
		\cdot
		\hat{\Delta}\left( \xi \right)
	\right]
	,
}$$
is a complete representation of an $N$-dimensional quantum state, defined as a kernel transformation from the density matrix ${\hat{\rho}}$ satisfying several key properties:
\begin{enumerate}
    \item
    It must be an informationally complete quasi-probability distribution on the hyper-sphere, real-valued with a self-conjugate kernel operator.
    \item
    Rotations of the spherical coordinates ${\xi}$ correspond to unitary transformations of the kernel ${\hat{\Delta}\left( \xi \right)}$.
\end{enumerate}
We reproduce the governing equations for these properties explicitly in Appendix~\ref{sec:the_stratonovich_weyl_conditions}.

Diagonalising a mixed-state density matrix, ${\hat{\rho}=\sum_{\psi}{p_{\psi}\lvert \psi \rangle\langle \psi \rvert}}$, the linearity of the trace allows us to similarly decompose the Wigner function as
\begin{align}
    W_{\hat{\rho}}
    \left( \xi \right)
	=
    \sum_{\psi}
    {
        p_{\psi}
        W_{\psi}
        \left( \xi \right)
    }
    .
\end{align}
If transitions between the states ${\lvert \psi \rangle\langle \psi \rvert}$ are in the group of kernel rotations defining our parameterisation, then the cyclic property of the trace allows us to further write
\begin{align}
    W_{\psi}
    \left( \xi \right)
    =
    W_{
        \psi_{0}
    }
    \left( R_{\psi} \cdot \xi_{0} \right)
\end{align}
with ${R_{\psi}}$ a coordinate rotation.
In such a case, the mixture ${\hat{\rho}}$ is represented in the spin-Wigner representation by a convex sum over rotations of the Wigner function of some reference state. We note that this is only possible for probabilistic-unitary noise channels. Where a noise process \emph{can} be represented as a convex combination of coordinate rotations, we can express it in terms of a convolution
\begin{align}
    \left(
        f \ast g
    \right)
    \left( x \right)
    =
    \int_{G}
    {
        f \left( y \right)
        g \left( y^{-1} x \right)
        \: dy
    }
    ,
\end{align}
where we take ${g\left(x\right)\rightarrow W_{\psi_{0}}\left(x\right)}$ and let the function ${f \left( y \right)}$ return the probability density for an erroneous coordinate rotation $y^{-1}$.

\subsection{Hyper-Spherical Harmonics}

Real, square-integrable functions ${f : S^{p-1} \rightarrow \mathbb{R}}$ on the ${\left(p-1\right)}$-sphere (including the spin Wigner functions) can be expressed as sums over spherical harmonics \cite{efthimiou_spherical_2014}:
\begin{align}
	f \left( \xi \right)
    =
    \sum^{\infty}_{n=0}
    {
        \sum^{N\left( p , n \right)}_{j=1}
        {
            c_{n,j} Y_{n,j} \left( \xi \right)
        }
    }
    .
\end{align}

A complete, orthonormal set of degree-${n}$ spherical harmonics has
\begin{align}
	N\left( p , n \right)
    =
    \frac{2n + p - 2}{n}
    \begin{pmatrix}
        n + p - 3 \\
        n - 1
    \end{pmatrix}
\end{align}
elements, and harmonics of distinct degrees are also orthogonal,
\begin{align}
    \int_{\xi}
	d\xi \:
	Y_{n,j}(\xi)
	\:
	Y_{m,k}(\xi)
	&=
	\delta_{n,m}
	\delta_{j,k}
	\nonumber\\
	\sum_{n,j}
	Y_{n,j}(\xi)
	\:
	Y_{n,j}(\eta)
	&=
	\delta
	\left(
		\xi
		-
		\eta
	\right)
	,
    \label{eq:harmonic_orthogonality_relations}
\end{align}
giving us the coefficient expression
\begin{align}
	c_{n,j}
    =
    \int_{\xi\in S^{p-1}}
    {
        f \left( \xi \right)
        \cdot
        Y_{n,j} \left( \xi \right)
        \: d\xi
    }
    .
\end{align}

Many functions will have a maximum degree $n_{max}$, the \emph{bandwidth}, above which all coefficients $c_{n,j}$ are zero. Following the convolution of two functions ${f}$ and ${g}$, we observe due to orthogonality that there can be no surviving terms for harmonics with degree greater than
\begin{align}
    n_{max}
    :=
    min
    \left(
        n^{\left(f\right)}_{max}
        ,
        n^{\left(g\right)}_{max}
    \right)
    .
\end{align}
As discussed in Appendix~\ref{sec:the_degree_wise_convolution_of_spherical_harmonics}, convolution may be treated independently for each degree $n$:
\begin{align}
    \left(
        f \ast g
    \right)
    \left( x \right)
    =
    \sum^{n_{max}}_{n=0}
    {
        \left(
            f_{n} \ast g_{n}
        \right)
        \left( x \right)
    }
    .
\end{align}
Further, in certain cases we will be interested in noise distributions that are invariant under rotations about some reference point $\xi_{0}$.
For functions of this type, the standard convolution theorem applies (also Appendix~\ref{sec:the_degree_wise_convolution_of_spherical_harmonics}), such that we may take the simple element-wise product over coefficients $c_{n,j}$. On $S^{2}$, this subclass of distributions corresponds to those harmonic expansions containing only \emph{zonal} harmonic functions.

Having introduced the key structures with which we will be working, we turn next to the question of how we can parameterise the spin Wigner function kernel, ${\hat{\Delta}\left( \xi \right)}$.

\section{Parameterisation via Harmonics}
\label{sec:parameterisation_via_harmonics_and_unitary_designs}

Prior to using the spin Wigner representation, we must choose a kernel ${\hat{\Delta}\left( \xi \right)}$ and its coordinates ${\xi}$. Accounting for normalisation and a global phase, any pure, {$N$-dimensional} quantum state undergoing quantum operations in SU$\left(N\right)$ is completely characterised by $2\left(N-1\right)$ angles, which contain all the information of the ${N-1}$ complex numbers in the standard state representation. It is however possible to construct informationally-complete Wigner functions using fewer than ${2\left(N-1\right)}$ parameters. For instance, the well-known Wigner function for the harmonic oscillator is capable of representing arbitrary rotations among the first $N$ Fock states, and yet is fully parameterised with only two variables that may be expressed in angles as ${q=\tanh{\left(\theta/2\right)}}$ and ${p=\tanh{\left(\phi/2\right)}}$.

In general, the fewer parameters in ${\xi}$, the more complex structures will appear in the Wigner function (i.e. the higher the degree of the maximum harmonic). As one potential choice, consider following the final example from \cite{tilma_wigner_2016} and choosing the kernel
\begin{align}
    \hat{\Delta}\left( \xi \right)
    &=
    \hat{U}_{N}\left(\xi\right)
    \left[
        I_{N}
        -
        \mathcal{N}\left(N\right)
        \hat{\Lambda}_{N^{2}-1}
    \right]
    \hat{U}^{\dagger}_{N}\left(\xi\right)
    \nonumber\\
    \mathcal{N}
    \left(N\right)
    &=
    \sqrt{\left(N+1\right)\left(N\right)\left(N-1\right)/2}
    ,
    \label{eq:tilma_full_lambda_kernel}
\end{align}
where ${\hat{\Lambda}_{N^{2}-1}}$ is the final, diagonal element of the generalised Gell-Mann matrices, under the standard construction.
Parameterising the spin-Wigner function by a maximal independent set of $2\left(N-1\right)$ angles ${\xi=\left\{\theta_{1}, \phi_{1}, \ldots, \theta_{N-1}, \phi_{N-1} \right\}}$, it then becomes equal (as noted in \cite{tilma_wigner_2016}) to the coherent-state based Wigner function of \cite{tilma_su_2012}, which simply quantifies the projection of the density matrix onto the spin coherent state ${\left\lvert \xi \right\rangle}$ \cite{nemoto_generalized_2000}. These spin coherent states therefore form the only structures of consequence.

While informationally complete, ${\left\lvert \xi \right\rangle}$ contains a number of parameters exponential in the qubit number, and an alternative qubit-local parameterisation with only ${2\log_{2}\left(N\right)}$ angles is described in \cite{rundle_simple_2017}. This alternative is more convenient for the study of qubit-local noise, and we build upon it in
Section~\ref{sub:tensor_product_structures}
and use it in several of the examples to follow. We would, however, like to be able to tailor our parameterisation to various other noise models, and so we now turn to identify more general requirements of a valid parameterisation.

\subsection{Requirements of Alternative Parameterisations}

We can construct valid kernels from any complete, orthogonal, Hermitian basis ${\{\hat{O}_{i}\}}$ for the traceless Hermitian matrices:
\begin{align}
    \hat{\Delta} \left( \xi \right)
	&=
	C_{\Delta} \mathds{1}
	+
	\sum_{i}
	\:
	\hat{O}_{i}
	\sum_{n,j}
	f^{(i)}_{(n,j)}
	Y_{n,j} \left( \xi \right)
	,
 \label{eq:general_kernel_form}
\end{align}
where ${f^{(i)}_{(n,j)}}$ is some complete set of discrete, orthogonal functions over the indices ${(n,j)}$.

There are three kernel requirements implied by the Stratonovich-Weyl relations, as summarised in Appendix~\ref{sec:the_stratonovich_weyl_conditions}.
Kernels in the form of Equation~\eqref{eq:general_kernel_form} satisfy the first requirement trivially, and the second by fixing ${C_{\Delta}}$ to achieve normalisation.
The third requirement depends on how we define the relationships between operators and spherical coordinates.
Since
\begin{align}
    \hat{\Delta} \left( R \xi \right)
	&=
	C_{\Delta} \mathds{1}
	+
	\sum_{n,j}
    \left(
        \sum_{i}
    	f^{(i)}_{(n,j)}
        \hat{O}_{i}
    \right)
	Y_{n,j} \left( R \xi \right)
    \nonumber\\
    &=
	C_{\Delta} \mathds{1}
	+
	\sum_{n,j}
    \left(
        \sum_{i}
    	f^{(i)}_{(n,j)}
        \hat{O}_{i}
    \right)
    \sum_{k}
    R^{(n)}_{j,k}
	Y_{n,k} \left( \xi \right)
    \nonumber\\
    &=
	C_{\Delta} \mathds{1}
	+
	\sum_{n,k}
    \left(
        \sum_{i}
        g^{(i)}_{(n,k)}
        \hat{O}_{i}
    \right)
    Y_{n,k} \left( \xi \right),
	\nonumber\\
    g^{(i)}_{(n,k)}
    &=
    \sum_{j} R^{(n)}_{j,k} f^{(i)}_{(n,j)}
    ,
\end{align}
where ${R^{(n)}}$ are orthogonal matrices and ${g^{(i)}_{(n,k)}}$ also form discrete orthogonal functions by the completeness of ${f^{(i)}_{(n,j)}}$.
A rotation of coordinates ${R}$ must therefore correspond to a unitary operation implementing the transformation
\begin{align}
    \sum_{i}
    f^{(i)}_{(n,j)}
    \hat{O}_{i}
    &\rightarrow
    \sum_{i}
    g^{(i)}_{(n,k)}
    \hat{O}_{i}
    .
\end{align}
For unspecified harmonic bases ${Y_{n,j}}$, the only way to guarantee this property is to support the entire orbit of the unitary transformation from the operator ${\hat{O}_{i}}$ on harmonics of the same degree ${n}$.
We note that the unitary operations corresponding to coordinate rotations need not explore the entire state space, and in general need not even form a group in their own right.

We next consider the structures of two known parameterisations, expressing them in the form of Equation~\eqref{eq:general_kernel_form} to show that they are equivalent.

\subsubsection{Example 1: Kernel as Displaced Parity Operator}

Our first example, introduced in \cite{tilma_wigner_2016} and described above in Equation~\eqref{eq:tilma_full_lambda_kernel}, has the kernel
\begin{align}
    \hat{\Delta}\left( \xi \right)
    &=
    \hat{U}_{N}\left(\xi\right)
    \left[
        I_{N}
        -
        \mathcal{N}\left(N\right)
        \hat{\Lambda}_{N^{2}-1}
    \right]
    \hat{U}^{\dagger}_{N}\left(\xi\right)
    \nonumber\\
    &=
    A
    I_{N}
    -
    B
    \left(
        \hat{U}_{N}\left(\xi\right)
        \lvert N \rangle
        \langle N \rvert
        \:
        \hat{U}^{\dagger}_{N}\left(\xi\right)
    \right)
    \nonumber\\
    &=
    A
    I_{N}
    -
    B
    \left\lvert \xi \right\rangle
    \left\langle \xi \right\rvert
    ,
\end{align}
with ${\left\lvert \xi \right\rangle}$ a spin coherent state and coefficients ${A,B}$ chosen to preserve normalisation.
Now, noting that \cite{tilma_su_2012}
\begin{align}
    \int d\xi \:
    \left\langle \xi \right\rvert
        \hat{\Lambda}_{i}
    \left\lvert \xi \right\rangle
    \left\langle \xi \right\rvert
        \hat{\Lambda}_{j}
    \left\lvert \xi \right\rangle
    &\propto
    \delta_{ij}
    ,
\end{align}
expanding the real, square-integrable function
${\left\langle \xi \right\rvert \hat{\Lambda}_{i} \left\lvert \xi \right\rangle}$
in harmonics we have
\begin{align}
    \left\langle \xi \right\rvert
        \hat{\Lambda}_{i}
    \left\lvert \xi \right\rangle
    &=
    \sum_{n,j}
    f^{(i)}_{(n,j)}
    Y_{n,j}(\xi)
\end{align}
with ${f^{(i)}_{(n,j)}}$ again drawn from some complete set of discrete orthogonal functions,
or
\begin{align}
    f^{(i)}_{(n,j)}
    &=
    \int d\xi\:
    Y_{n,j}(\xi)
    \left\langle \xi \right\rvert
        \hat{\Lambda}_{i}
    \left\lvert \xi \right\rangle
    \nonumber\\
    &=
    \mathrm{Tr}
    \left[
        \hat{D}_{n,j}
        \hat{\Lambda}_{i}
    \right]
\end{align}
with ${\hat{D}_{n,j}}$ defined to be
\begin{align}
    \hat{D}_{n,j}
    &=
    \int d\xi\:
    Y_{n,j}(\xi)
    \lvert \xi \rangle \langle \xi \rvert
    .
\end{align}

The kernel is now
\begin{align}
    \hat{\Delta}\left( \xi \right)
    &=
    A
    I_{N}
    -
    B
    \lvert \xi \rangle
    \langle \xi \rvert
    \nonumber\\
    &=
    A
    I_{N}
    -
    B
    \sum_{i}
    \left\langle \xi \right\rvert
        \hat{\Lambda}_{i}
    \left\lvert \xi \right\rangle
    \hat{\Lambda}_{i}
    \nonumber\\
    &=
    A
    I_{N}
    -
    B
    \sum_{i}
    \hat{\Lambda}_{i}
    \sum_{n,j}
    f^{(i)}_{n,j}
    Y_{n,j}(\xi)
    ,
\end{align}
which, with the identification
${\hat{\Lambda}_{i}\rightarrow \hat{O}_{i}}$
takes the form of Equation~\eqref{eq:general_kernel_form}.

\subsubsection{Example 2: Generalised Displacement Operators}
\label{subs:describe_brif_method}

In \cite{brif_phase-space_1999} the authors introduced a kernel construction
based on harmonic-weighted spin coherent states:
\begin{align}
    \hat{\Delta}(\xi)
	=
	C_{\Delta}
	\sum_{n,j}
	Y_{n,j}(\xi)
	\hat{D}_{n,j}
	,
\end{align}
with normalising constant ${C_{\Delta}}$ and operator basis
\begin{align}
    \hat{D}_{n,j}
	=
	\int_{X} d\xi
	\:
	Y_{n,j}(\xi)
	\:
	\lvert \xi \rangle \langle \xi \lvert
    .
    \label{eq:coherent_state_generalised_displacement}
\end{align}
We briefly recap why this construction satisfies the Stratonovich--Weyl conditions in Appendix~\ref{sec:the_harmonic_displacement_kernel_construction}.

Because we have chosen to use real-valued spherical harmonics, the generalised displacement operators
used as an operator basis here are Hermitian, and can therefore be expanded in any of the common bases for Hermitian operators. Choosing the generalised Gell-Mann matrices gives us directly
\begin{align}
    \hat{\Delta}(\xi)
	=
	C_{\Delta}
	+
    \sum_{i}
    \hat{\Lambda}_{i}
    \sum_{n,j}
    \frac{1}{2}
    \mathrm{Tr}
    \left[
        \hat{D}_{n,j}
        \hat{\Lambda}_{i}
    \right]
	Y_{n,j}(\xi)
	,
\end{align}
which has the form we require on taking the substitution
\begin{align}
    \frac{1}{2}
    \mathrm{Tr}
    \left[
        \hat{D}_{n,j}
        \hat{\Lambda}_{i}
    \right]
    &\rightarrow
    f^{(i)}_{(n,j)}
    .
\end{align}
The orthogonality of the functions ${f^{(i)}_{(n,j)}}$
can be verified by exploiting the second orthogonality relation
for the spherical harmonics in Equation~\eqref{eq:harmonic_orthogonality_relations}:
\begin{align}
    \sum_{n,j}
    \mathrm{Tr}
    \left[
        \hat{D}_{n,j}
        \hat{\Lambda}_{i}
    \right]
    \mathrm{Tr}
    \left[
        \hat{D}_{n,j}
        \hat{\Lambda}_{k}
    \right]
    &=
    \sum_{n,j}
    \int d\xi
	\:
	Y_{n,j}(\xi)
	\:
	\langle \xi \rvert \hat{\Lambda}_{i} \lvert \xi \rangle
    \nonumber\\
    &\qquad \cdot
    \int d\eta
	\:
	Y_{n,j}(\eta)
	\:
	\langle \eta \rvert \hat{\Lambda}_{k} \lvert \eta \rangle
    \nonumber\\
    &=
    \int d\xi
	\:
	\langle \xi \rvert \hat{\Lambda}_{i} \lvert \xi \rangle
	\langle \xi \rvert \hat{\Lambda}_{k} \lvert \xi \rangle
    \nonumber\\
    &\propto
    \delta_{ik}
    .
\end{align}

As described in Appendix~\ref{sec:the_harmonic_displacement_kernel_construction},
this construction (and therefore also the previous one) avoids the problem of choosing functions ${f^{(i)}_{(n,j)}}$
consistently across harmonics of different degrees by defining the coordinates
of the harmonic functions and the spin coherent states to be co-variant in the
definition of ${\hat{D}_{n,j}}$. This reduces coordinate transformations at all degrees
to a change of coordinate for a single coherent state ${\left\lvert \xi \right\rangle}$.
As we will see in the sections to follow, when transformations of the kernel do not span the
full quantum state space these co-variant states need not span the full space of
${\mathrm{SU}(N)}$ spin coherent states.

\subsection{Tensor Product Structures}
\label{sub:tensor_product_structures}

When we come to consider noise models in Section~\ref{sec:example_noise_models},
we will need to be able to define coefficients ${f^{(i)}_{(n,j)}}$ consistent with
unitary transformations \emph{restricted to operations induced by the noise itself}.
In many cases, this noise will operate locally on some subsystem, and in such cases we
can induce a tensor product structure on the kernel that allows us to factorise
${f^{(i)}_{(n,j)}}$ and solve for the factors independently in spaces of smaller dimension.

Take as an example the qubit-local parameterisation of \cite{rundle_simple_2017}:
\begin{align}
    \hat{\Delta}(\xi)
    &=
    \hat{U}(\xi)
    \hat{\Pi}
    \hat{U}^{\dagger}(\xi)
    \nonumber\\
    \hat{U}(\xi)
    &\in
    \mathrm{SU}(2)^{\otimes m}
    \nonumber\\
    \hat{\Pi}
    &=
    I_{N}
    -
    \mathcal{N}\left(N\right)
    \hat{\Lambda}_{N^{2}-1}
    .
\end{align}
This can be generalised straightforwardly to subsystems of any dimension with
\begin{align}
    \hat{U}(\xi)
    &\in
    \bigotimes^{m}_{k=1}
    \mathrm{SU}(N_{i})
    \nonumber\\
    \hat{\Pi}
    &=
    I_{N}
    -
    \mathcal{N}\left(N\right)
    \hat{\Lambda}_{N^{2}-1}
    \nonumber\\
    &=
    A I_{N}
    -
    B
    \lvert N \rangle \langle N \rvert
    \nonumber\\
    &=
    A I_{N}
    -
    B
    \bigotimes^{m}_{k=1}
    \lvert N_{k} \rangle \langle N_{k} \rvert_{k}
    ,
\end{align}
where again coefficients ${A}$ and ${B}$ are chosen to preserve normalisation,
and ${\lvert N \rangle}$ is the lowest-weight state.

Distributing ${\hat{U}(\xi)}$ across the tensor product structure of the separable state
${\lvert N \rangle \langle N \rvert}$ gives
\begin{align}
    \hat{U}(\xi)
    \lvert N \rangle \langle N \rvert
    \hat{U}^{\dagger}(\xi)
    &=
    \bigotimes^{m}_{k=1}
    \hat{U}_{k}(\xi_{i})
    \lvert N_{k} \rangle \langle N_{k} \rvert_{k}
    \hat{U}^{\dagger}_{k}(\xi_{k})
    \nonumber\\
    &=
    \bigotimes^{m}_{k=1}
    \lvert \xi_{k} \rangle \langle \xi_{k} \rvert_{k}
    ,
\end{align}
with ${\lvert \xi_{k} \rangle}$ a spin coherent state for subsystem ${k}$.
Expanding these states in lambda-matrices ${\Lambda^{(k)}_{i_{k}}}$ for each individual subsystem ${k}$
and extending the index for the operator basis ${i \rightarrow i_{1}, \ldots, i_{m}}$,
we then have
\begin{align}
    f^{(i)}_{(n,j)}
    &\rightarrow
    \prod^{m}_{k=1}
    f^{(i_{k})}_{(n,j)}
    \nonumber\\
    f^{(i_{k})}_{(n,j)}
    &=
    \mathrm{Tr}
    \left[
        \hat{D}^{(k)}_{n,j}
        \hat{\Lambda}^{(k)}_{i_{k}}
    \right]
    \nonumber\\
    \hat{D}^{(k)}_{n,j}
    &=
    \int d\xi_{k}
	\:
	Y^{(k)}_{n,j}(\xi_{k})
	\:
	\lvert \xi_{k} \rangle
	\langle \xi_{k} \rvert_{k}
    ,
\end{align}
where harmonics ${Y^{(k)}_{n,j}}$ have been labelled with the subsystem
index to indicate that they are defined over the restricted spherical space
defined by coordinates ${\xi_{k}}$

The kernel is now
\begin{align}
    \hat{\Delta}\left( \xi \right)
    &=
    A
    I_{N}
    -
    B
    \bigotimes^{m}_{k=1}
    \lvert \xi_{k} \rangle \langle \xi_{k} \rvert_{k}
    \nonumber\\
    &=
    A
    I_{N}
    -
    B
    \bigotimes^{m}_{k=1}
    \sum_{i_{k}}
        \left\langle \xi_{k} \right\rvert
            \hat{\Lambda}^{(k)}_{i_{k}}
        \left\lvert \xi_{k} \right\rangle
        \hat{\Lambda}^{(k)}_{i_{k}}
    \nonumber\\
    &=
    A
    I_{N}
    -
    B
    \sum_{i_{1},\ldots,i_{m}}
    \prod^{m}_{k=1}
        \left\langle \xi_{k} \right\rvert
            \hat{\Lambda}^{(k)}_{i_{k}}
        \left\lvert \xi_{k} \right\rangle
    \bigotimes^{m}_{k=1}
        \hat{\Lambda}^{(k)}_{i_{k}}
    \nonumber\\
    &=
    A
    I_{N}
    -
    B
    \sum_{i_{1},\ldots,i_{m}}
    \sum_{n,j}
        \prod^{m}_{k=1}
        f^{(i_{k})}_{(n,j)}
        Y^{(k)}_{n,j}(\xi_{k})
        \bigotimes^{m}_{k=1}
        \hat{\Lambda}^{(k)}_{i_{k}}
    ,
\end{align}
which, with the identifications
\begin{align}
    \prod^{m}_{k=1}
    f^{(i_{k})}_{(n,j)}
    &\rightarrow
    f^{(i)}_{(n,j)}
    \nonumber\\
    \prod^{m}_{k=1}
    Y^{(k)}_{n,j}(\xi_{k})
    &\rightarrow
    Y_{n,j}(\xi)
    \nonumber\\
    \bigotimes^{m}_{k=1}
    \hat{\Lambda}^{(k)}_{i_{k}}
    &\rightarrow
    \hat{O}_{i}
    ,
\end{align}
takes the form of Equation~\eqref{eq:general_kernel_form}
(harmonics not decomposing in this product structure have
zero-coefficients).

\section{Example Noise Models}
\label{sec:example_noise_models}

We have now seen how the spin Wigner function may be parameterised using different operator bases and tensor product structures.
In this section, we turn to consider several noise models and construct kernels to efficiently represent their influence on the quantum state.

\subsection{Noise in a Lie Subgroup}
\label{sub:noise_in_a_lie_subgroup}

Suppose that operations generated by noise exist in a Lie subgroup (say, ${H}$) of ${\mathrm{U}(N)}$,
generated by an algebra over Hermitian elements ${\{\hat{H}_{k}\}}$ such that
\begin{align}
    \exists a_{k} \in \mathbb{R}
    \quad
    s.t.
    \quad
    h
    &=
    e^{i\left( \sum_{k} a_{k} \hat{H}_{k} \right)}
    ,\quad
    \forall h \in  H
    .
\end{align}
This algebra is clearly closed under Hermitian conjugation and so, borrowing
from representation theory as applied in quantum error correction \cite{knill_theory_2000},
the operator Hilbert space is isomorphic to
\begin{align}
    \mathcal{H}
    &\cong
    \bigoplus^{K}_{k=1}(A_{k}\otimes B_{k})
    ,
    \label{eq:tensor_sum_algebra_decomposition}
\end{align}
such that for all ${k}$, ${\mathcal{L}(A_{k})}$ forms an irreducible representation for the action of
the algebra on ${A_{k}}$, the action on ${B_{k}}$ corresponds to the identity,
and
\begin{align}
    \sum_{k}
        \mathrm{dim}(A_{k})
        \mathrm{dim}(B_{k})
    &=
    \mathrm{dim}(\mathcal{H})
    .
\end{align}

Now, let ${\{\hat{a}_{k,i}\}}$ and ${\{\hat{b}_{k,j}\}}$ be bases of traceless Hermitian operators over ${A_{k}}$ and ${B_{k}}$
respectively, ${\{\hat{c}^{(m)}_{kl}\}}$ the basis elements over the full Hilbert space that allow transitions between the ${k}$ and ${l}$ sub-spaces,
and ${\{\hat{d}_{k}\}}$ the basis elements inducing a relative phase between subspace ${k}$ and all others.
These operators give us a basis in which to expand the kernel.
As described in Appendix~\ref{sec:app_07_noise_restricted_kernel}, decompose this expansion
in noise and residual coordinates ${\bar{\xi}}$ and ${\bar{\eta}}$ respectively.
This gives
\begin{widetext}
    \begin{align}
        \hat{\Delta}(\bar{\xi},\bar{\eta})
        &=
        C_{\Delta}
        \mathbb{I}
        +
        \sum^{K}_{k=1}
        \sum_{j}
        \left(
            \sum_{i}
            \hat{a}_{k,i}
            \sum_{n,l}
            \left[
                \sum_{n',p}
                q^{(k,i,j)}_{(n,l,n',p)}
                Y_{n',p}(\bar{\eta})
            \right]
            Y_{n,l}(\bar{\xi})
        \right)
        \otimes
        \hat{b}_{k,j}
        \nonumber\\
        &\quad
        +
        \sum^{K}_{k=1}
        \sum^{K}_{l=k+1}
        \sum^{\mathrm{dim}(B_{k})}_{k'=1}
        \sum^{\mathrm{dim}(B_{l})}_{l'=1}
        \left(
            \sum^{2 \mathrm{dim}(A_{k}) \mathrm{dim}(A_{l})}_{m=1}
            \hat{c}^{(m)}_{klk'l'}
            \sum_{n,j}
            \left[
                \sum_{n',p}
                q'^{(k,l,k',l')}_{(n,j,n',p)}
                Y_{n',p}(\bar{\eta})
            \right]
            Y_{n,j}(\bar{\xi})
        \right)
        \nonumber\\
        &\quad
        +
        \sum^{K}_{k=2}
        \mathcal{Q}_{\hat{d}_{k}}(\bar{\eta})
        \hat{d}_{k}
        .
        \label{eq:algebra_kernel_general}
    \end{align}
\end{widetext}
Here ${\mathcal{Q}_{\hat{d}_{k}}(\bar{\eta})}$ are orthogonal functions in the residual coordinates ${\bar{\eta}}$,
since the operators ${\hat{d}_{k}}$ are invariant under the action of the noise algebra.

Though Equation~\ref{eq:algebra_kernel_general} appears quite unwieldy in its current form, an explicit, orthogonal expansion in terms of harmonics ${Y_{n,j}(\bar{\xi})}$ is required only within groups of operators equivalent up to the action of the noise-algebra.
Further, in practice we will only reference the coordinates ${\bar{\eta}}$
when integrating over the full space.
These observations allow the kernel to be replaced for our purposes with the simpler effective expression
\begin{widetext}
    \begin{align}
        \hat{\Delta}(\bar{\xi},\eta)
        &=
        C_{\Delta}
        \mathbb{I}
        +
        \sum^{K}_{k=1}
        \sum_{j}
        \sin( \omega_{k,j} \eta )
        \left(
            \sum_{i}
            \hat{a}_{k,i}
            \sum_{n,j}
            f^{(k,i)}_{(n,l)}
            Y_{n,l}(\bar{\xi})
        \right)
        \otimes
        \hat{b}_{k,j}
        \nonumber\\
        &\quad
        +
        \sum^{K}_{k=1}
        \sum^{K}_{l=k+1}
        \sum^{\mathrm{dim}(B_{k})}_{k'=1}
        \sum^{\mathrm{dim}(B_{l})}_{l'=1}
        \sin( \Omega_{k,l,k',l'} \eta )
        \left(
            \sum^{2 \mathrm{dim}(A_{k}) \mathrm{dim}(A_{l})}_{m=1}
            \hat{c}^{(m)}_{klk'l'}
            \sum_{n,j}
            f'^{(k,l)}_{(n,j)}
            Y_{n,j}(\bar{\xi})
        \right)
        \nonumber\\
        &\quad
        +
        \sum^{K-1}_{k=1}
        \cos( k \eta)
        \hat{d}_{k+1}
        ,
        \label{eq:algebra_kernel_effective}
    \end{align}
\end{widetext}
where
\begin{align}
    \omega_{k,j}
    &=
    \mathrm{lex}(k,j)
    \nonumber\\
    \Omega_{k,l,k',l'}
    &=
    \sum^{K}_{k=1} \left( \mathrm{dim}(B_{k})^{2} - 1 \right)
    + \mathrm{lex}(k,l,k',l')
    ,
\end{align}
and ${\mathrm{lex}(i,j,k,\ldots)}$ returns the position in any lexicographic order over the argument-indices.
To satisfy the orthogonality requirements of the coefficients ${f^{(k,i)}_{(n,l)}}$, we may apply the generalised
displacement operators of \cite{brif_phase-space_1999} (described in Equation~\eqref{eq:coherent_state_generalised_displacement}), but now with states defined in the
reduced space of noise-induced rotations.

\subsubsection{Example 1: Single Qubit Dephasing}
\label{sec:single_qubit_dephasing}

As an example, consider single-qubit dephasing.
In this case, the noise may be expressed as a classical distribution over evolution operators
\begin{align}
    \hat{U}(\theta)
    &=
    e^{i\theta\hat{\sigma}_{z}}
    ,
\end{align}
which clearly form a Lie subgroup.
The Hilbert space decomposes as the tensor-sum of two one-dimensional
spaces (the eigenstates of ${\hat{\sigma}_{z}}$),
and the effective kernel becomes
\begin{align}
    \hat{\Delta}(\theta,\eta)
    &=
    C_{\Delta}
    \mathbb{I}
    +
    \cos( \eta )
    \hat{\sigma}_{z}
    \nonumber\\
    &\quad
    +
    \sin( 3 \eta )
    \left(
        \hat{\sigma}_{x}
        \cos(2\theta)
        +
        \hat{\sigma}_{y}
        \sin(2\theta)
    \right)
    ,
\end{align}
where the coefficients are derived from the circular space of states
${\lvert \theta \rangle = ( e^{-i\theta} , e^{i\theta} )^{T}}$
and harmonics
${\sin(n\theta)}$, ${\cos(n\theta)}$
as
\begin{align}
    \sum_{n,j}
    \mathrm{Tr}
    \left[
        \hat{D}_{n,j}
        \hat{\sigma}_{x}
    \right]
    Y_{n,j}(\theta)
    &\propto
    \int d\phi \:
        \delta(\theta - \phi)
    \langle \phi \rvert
    \hat{\sigma}_{x}
    \lvert \phi \rangle
    \nonumber\\
    &=
    \langle \theta \rvert
    \hat{\sigma}_{x}
    \lvert \theta \rangle
    \nonumber\\
    &=
    \cos(2\theta)
    .
\end{align}
In analogy with the properties of the spherical harmonics and spin coherent states, we observe that
\begin{align}
    \sum^{\infty}_{n=0} \sin(n\phi) \sin(n\theta) + \cos(n\phi) \cos(n\theta)
    &=
    \pi
    \delta ( \phi - \theta )
    +
    \frac{1}{2}
    ,\nonumber\\
    \int d\theta
    \lvert \theta \rangle
    \langle \theta \rvert
    &\propto
    \mathbb{I}
    ,\nonumber\\
    \int d\theta
    \langle \theta \rvert
        \hat{\sigma}_{j}
    \lvert \theta \rangle
    \langle \theta \rvert
        \hat{\sigma}_{k}
    \lvert \theta \rangle
    &=
    \begin{cases}
        \delta_{jk} & j \neq 3 \\
        0 & j = 3
    \end{cases}
    .
\end{align}
The anomalous case
${\langle \theta \rvert \hat{\sigma}_{z} \lvert \theta \rangle=0}$
does not cause an issue here, as we are only interested in generating
orthogonal coefficients for ${\hat{\sigma}_{x}}$ and ${\hat{\sigma}_{y}}$.

We make the identifications
\begin{align}
    \hat{\sigma}_{z}
    &\rightarrow
    \hat{d}_{2}
    ,\quad
    \hat{\sigma}_{x}
    \rightarrow
    \hat{c}^{(1)}_{1,2}
    ,\quad
    \hat{\sigma}_{y}
    \rightarrow
    \hat{c}^{(2)}_{1,2}
    .
\end{align}
There are no explicit
${\hat{a}_{k,i}}$ or ${\hat{b}_{k,j}}$ operators in this expression.
This is a common feature when the noise forms a maximally commuting subgroup,
since all generators may be simultaneously diagonalised
and the structure collapses to
\begin{align}
    \mathrm{dim}(A_{k})
    =
    \mathrm{dim}(B_{k})
    =
    1
    .
\end{align}

\subsubsection{Example 2: Two-Qubit Rotations}

Our next example is the set of rotations generated by the ${\hat{\sigma}^{(1)}_{z}\hat{\sigma}^{(2)}_{z}}$ operator,
\begin{align}
    \hat{U}(\theta)
    &=
    e^{i \theta \hat{\sigma}^{(1)}_{z} \hat{\sigma}^{(2)}_{z} }
    .
\end{align}
In this case, we do not have a maximal commuting subgroup, so there will be a non-trivial tensor-product structure,
\begin{align}
    \mathrm{dim}(A_{k}) = 1
    ,\quad
    \mathrm{dim}(B_{k}) = 2
    ,\quad
    k \in \{1, 2\}
    .
\end{align}

We may choose
\begin{align}
    \hat{b}_{k,1}
    &=
    \hat{\sigma}^{(1)}_{z}
    \hat{P}_{A}(k)
    \nonumber\\
    \hat{b}_{k,2}
    &=
    \hat{\sigma}^{(1)}_{x} \hat{\sigma}^{(2)}_{x}
    \hat{P}_{A}(k)
    \nonumber\\
    \hat{b}_{k,3}
    &=
    \hat{\sigma}^{(1)}_{y} \hat{\sigma}^{(2)}_{x}
    \hat{P}_{A}(k)
    \nonumber\\
    \hat{P}_{A}(k)
    &=
    \left( \mathbb{I} + (-1)^{k} \hat{\sigma}^{(1)}_{z} \hat{\sigma}^{(2)}_{z} \right)
    ,
\end{align}
and the residual operators
\begin{align}
    \hat{c}^{(1)}_{1,2,m,n}
    &=
    \left( \hat{\sigma}^{(1)}_{x} + \hat{\sigma}^{(2)}_{x} \right)
    \left[
        \hat{P}_{B}(m)\hat{P}_{A}(1) + \hat{P}_{B}(n)\hat{P}_{A}(2)
    \right]
    \nonumber\\
    \hat{c}^{(2)}_{1,2,m,n}
    &=
    \left( \hat{\sigma}^{(1)}_{y} + \hat{\sigma}^{(2)}_{y} \right)
    \left[
        \hat{P}_{B}(m)\hat{P}_{A}(1) + \hat{P}_{B}(n)\hat{P}_{A}(2)
    \right]
    \nonumber\\
    \hat{d}_{2}
    &=
    \hat{\sigma}^{(1)}_{z}
    \hat{\sigma}^{(2)}_{z}
    \nonumber\\
    \hat{P}_{B}(m)
    &=
    \left( \mathbb{I} + (-1)^{m} \hat{\sigma}^{(1)}_{z} \right)
    ,
\end{align}
where the ${\hat{c}^{(1/2)}_{1,2,m,n}}$ are related through ${\hat{U}(\theta)}$ according to
\begin{align}
    \hat{c}^{(1)}_{1,2,m,n}
    \cos(2\theta)
    +
    \hat{c}^{(2)}_{1,2,m,n}
    \sin(2\theta)
\end{align}
as in the previous example.

\subsubsection{Example 3: Charge-preserving Transformations}
\label{subs:charge_preserving_transformations}

Next we consider the group of transformations generated by operators of the form
\begin{align}
    \hat{U}( \phi_{1}, \phi_{2}, \theta )
    &=
    e^{i \phi_{1} \hat{J}^{(1)}_{z} }
    e^{i \phi_{2} \hat{J}^{(2)}_{z} }
    e^{i \theta \left( \hat{J}^{(1)}_{+} \hat{J}^{(2)}_{-} + \hat{J}^{(1)}_{-} \hat{J}^{(2)}_{+} \right) }
    .
    \label{eq:example_3_charge_preserving_transformations}
\end{align}
Since the factor terms in ${\hat{U}}$ no longer commute, we no longer have that ${\mathrm{dim}(A_{k})=1}$.
The conservation of charge implied by the Jaynes--Cummings-like operator ${\hat{J}^{(1)}_{+} \hat{J}^{(2)}_{-} + \hat{J}^{(1)}_{-} \hat{J}^{(2)}_{+}}$ nonetheless imposes a block-diagonal structure on the state space.
Denoting the dimensions of the first and second subsytem ${d_{1}}$ and ${d_{2}}$ respectively, and the charge number ${J}$, the subspace ${A_{J}}$ has dimension
\begin{align}
    D_{J}
    &=
    \begin{cases}
        J+1 & J \leq d_{1},d_{2} \\
        d_{1}+1 & d_{1} \leq J \leq d_{2} \\
        d_{2}+1 & d_{2} \leq J \leq d_{1} \\
        d_{1} + d_{2} - J + 1 & d_{1},d_{2} \leq J
    \end{cases}
\end{align}

Since ${\hat{J}^{(1/2)}_{\pm}}$ act as ladder operators for a one-dimensional chain of states,
in the absence of ${\hat{J}^{(1/2)}_{z}}$ the
subspaces ${A_{J}}$ with the same dimension could be identified.
The multiplicity of dimension ${D_{J}}$ would then define the dimension of a corresponding invariant tensor
product space ${B_{D_{J}}}$.
However, the inclusion of operators ${\hat{J}^{(1/2)}_{z}}$ breaks this additional symmetry.

The charge is preserved within each subspace, so that the action of
${e^{i\phi(\hat{J}^{(1)}_{z} + \hat{J}^{(2)}_{z})}}$
on subspace ${A_{J}}$ is to add a global phase
${e^{i\phi J}}$.
We can therefore make the substitution
${ e^{i\phi_{2} \hat{J}^{(2)}_{z} } \rightarrow e^{i\phi_{2} J} e^{-i\phi_{2} \hat{J}^{(1)}_{z} } }$.
Next, observing that ${\hat{J}^{(1)}_{z}}$ and
${\hat{J}^{(1)}_{+} \hat{J}^{(2)}_{-} + \hat{J}^{(1)}_{-} \hat{J}^{(2)}_{+}}$
generate the group ${\mathrm{SU}(2)}$ in ${A_{J}}$,
within this subspace we use the ${\mathrm{SU}(2)}$ spin coherent states in ${D_{J}}$ dimensions
\cite{nemoto_generalized_2000},
\begin{align}
    \lvert \phi_{1}, \phi_{2}, \theta \rangle
    &=
    \sum^{D_{J}-1}_{j=0}
    e^{ij\phi_{2}}
    e^{i(D_{J}-j-1)\phi_{1}}
    \sin^{j}(\theta)
    \cos^{D_{J}-j-1}(\theta)
    \nonumber\\
    &\qquad \qquad \cdot
    \begin{pmatrix}
        D_{J}-1 \\ j
    \end{pmatrix}^{1/2}
    \lvert D_{J} - j -1, j \rangle
    .
\end{align}
The total state space is then generated by the action of ${\hat{U}}$ on
the highest-weight composite state
\begin{align}
    \lvert \xi_{0} \rangle
    &=
    \bigoplus_{A_{J}}
    \lvert D_{J} , 0 \rangle
    .
    \label{eq:tensor_sum_su2_spin_coh_states}
\end{align}
The special cases ${J\in\{0,d_{1}+d_{2}\}}$ span the kernel of the exchange operator, and are left invariant up to respective global phases.

Although the ${\mathrm{SU}(2)}$ spin coherent states in each subspace give respective resolutions of unity on integration,
we cannot apply the method of \cite{brif_phase-space_1999} to the full space as described in Section~\ref{subs:describe_brif_method}.
Constructing a pseudo-displacement operator
\begin{align}
    \hat{D}_{n,j}
    &=
    \int d\xi
    Y_{n,j} ( \xi )
    \lvert \xi \rangle
    \langle \xi \rvert
    ,
    \nonumber\\
    \lvert \xi \rangle
    &=
    \hat{U}( \phi_{1}, \phi_{2}, \theta )
    \lvert \xi_{0} \rangle
    ,
\end{align}
does not allow us to construct orthogonal coefficients for all
elements of a complete basis of traceless Hermitian operators in ${N=d_{1}\cdot d_{2}}$ dimensions.
Nonetheless, we may identify subsets of basis operators for which orthogonal coefficients
can be constructed.
These correspond to the parenthetic sums in Equation~\eqref{eq:algebra_kernel_effective},
and include for instance the X- and Y-type lambda matrices for a given transition,
which are related to one another through the phase rotations
${e^{i \phi_{1} \hat{J}^{(1)}_{z} }}$ and ${e^{i \phi_{2} \hat{J}^{(2)}_{z} }}$.
Additionally, since we can have ${\mathrm{dim}(A_{J})>1}$, there is now many-to-one relationship between the irreducible subspaces ${A_{k}}$ for the state space and for the basis operators.
The particular orthogonality requirements for this construction are
discussed further in Appendix~\ref{sec:orth}.

\subsection{Noise Inducing only Limited Entanglement}

When considering noisy gates that can generate small amounts of entanglement,
consider the algebra of entangling operations generated in the small-angle limit,
\begin{align}
    e^{i\theta_{12}\hat{E}_{12}}
    e^{i\theta_{23}\hat{E}_{23}}
    \approx
    \mathbb{I}
    +
    i\theta_{12}\hat{E}_{12}
    +
    i\theta_{23}\hat{E}_{23}
    +
    \mathcal{O}(\theta^{2})
    .
\end{align}
In this regime, and allowing conjugation by arbitrary local rotations,
we now have operators of the form
\begin{align}
    &
    \hat{U}( \bar{\eta}_{1,2,3}, \bar{\phi}_{1,2,3}, \theta_{12}, \theta_{23} )
    =
    \nonumber\\
    &\quad
    \prod^{3}_{j=1}
    e^{i \bar{\phi}_{j} \cdot \bar{\sigma}^{(j)} }
    \left[
        \mathbb{I}
        +
        i \theta_{12} \hat{C}^{(12)}_{X}
        +
        i \theta_{23} \hat{C}^{(23)}_{X}
    \right]
    \prod^{3}_{j=1}
    e^{i \bar{\eta}_{j} \cdot \bar{\sigma}^{(j)} }
    ,
    \label{eq:weak_entangling_noise_example}
\end{align}
where ${\hat{C}^{(kj)}_{X}}$ is the controlled-NOT gate from qubit ${k}$ to qubit ${j}$.
These operators are not closed under composition and so no longer form a group.
This is a necessary restriction as arbitrary quantum circuits can be generated
from the composition of one- and two-qubit gates.
This section is the first where we have had to restrict the set of transformations so that they do not form a group.

Following the method of Vidal \cite{vidal_efficient_2003},
consider a quantum state expressed recursively in the Schmidt basis
\begin{align}
    \left\lvert
        \Psi
    \right\rangle
    &=
    \sum^{\chi}_{\alpha_{1}=1}
    \lambda^{[1]}_{\alpha_{1}}
    \left\lvert
        \Phi^{[1]}_{\alpha_{1}}
    \right\rangle
    \left\lvert
        \Phi^{[2,3]}_{\alpha_{1}}
    \right\rangle
    \nonumber\\
    &=
    \sum^{1}_{i_{1,2,3}=0}
    \sum^{\chi}_{\alpha_{1,2,3}=1}
    \Gamma^{[1]i_{1}}_{\alpha_{1}}
    \lambda^{[1]}_{\alpha_{1}}
    \Gamma^{[2]i_{2}}_{\alpha_{1}\alpha_{2}}
    \lambda^{[2]}_{\alpha_{2}}
    \Gamma^{[3]i_{3}}_{\alpha_{2}}
    \lvert
        i_{1}
        i_{2}
        i_{3}
    \rangle
    .
\end{align}
Here ${\chi=2}$ is the maximal Schmidt rank chosen to restrict the space of allowable states,
${\lambda^{[j]}_{\alpha_{j}}}$ are the Schmidt coefficients for the ${j}$th round of decomposition,
and ${\Gamma^{[j]i_{j}}_{\alpha_{k}\alpha_{j}}}$
are the coefficients of the computational basis states for qubit ${j}$ within
the Schmidt basis vector of index ${(\alpha_{k},\alpha_{j})}$.
Before taking the sum over ${\alpha_{j}}$, the tensor
\begin{align}
    \Gamma^{[1]i_{1}}_{\alpha_{1}}
    \lambda^{[1]}_{\alpha_{1}}
    \Gamma^{[2]i_{2}}_{\alpha_{1}\alpha_{2}}
    \lambda^{[2]}_{\alpha_{2}}
    \Gamma^{[3]i_{3}}_{\alpha_{2}}
\end{align}
provides a convenient representation for studying the action of
${\hat{U}( \bar{\eta}_{1,2,3}, \bar{\phi}_{1,2,3}, \theta_{12}, \theta_{23} )}$
on the state space.
Vectorising this tensor, we observe that the
Schmidt coefficient pair ${\lambda^{[j]}_{\alpha_{j}\in\{0,1\}}}$ (for each ${j}$)
is equivalent to introducing an entangled pair
\begin{align}
    \lambda^{[j]}_{0}
    \lvert 00 \rangle
    +
    \lambda^{[j]}_{1}
    \lvert 11 \rangle
\end{align}
across the ${j}$th partition,
and that the sum over ${\alpha_{j}}$ is equivalent to projecting this pair
onto the fixed state of zero relative phase in the Fourier basis.

Equation~\ref{eq:weak_entangling_noise_example} contains only CNOT entangling gates,
which may be generated with local operations and the consumption of a maximally-entangled state
${\lvert\Phi^{+}\rangle=(\lvert 00 \rangle + \lvert 11 \rangle)/\sqrt{2}}$.
Let the density matrix corresponding to this Bell state be denoted ${\hat{\rho}_{\Phi^{+}}}$.
Transform an initial ${n}$-qubit density matrix ${\hat{\rho}}$ according to
\begin{align}
    \hat{\rho}
    &\rightarrow
    \hat{\rho}'
    =
    \hat{\rho}
    \otimes
    \hat{\rho}^{\otimes(n-1)}_{\Phi^{+}}
    ,
\end{align}
and observables of interest according to
\begin{align}
    \hat{O}
    &\rightarrow
    \hat{O}'
    =
    \hat{O}
    \otimes
    \lvert ++ \rangle\langle ++ \rvert^{\otimes(n-1)}_{\Phi^{+}}
    .
\end{align}
The action of the noise is now completely described within a tensor product
coordinate representation of the kind described in Section~\ref{sub:tensor_product_structures},
where the local qudit systems are composed of a single computational
qubit alongside one ancillary qubit from each of up to two Bell pairs.
The dimension of this coordinate representation continues to scale linearly with the number
of computational qubits, but also increases now with the maximum allowable Schmidt rank.

\section{Error-Mitigation via Rescaled Expectation values}
\label{sec:error_mitigation_via_rescaled_expectation_values}

The space of observable expectation values, their \emph{joint numerical range} \cite{xu_bounding_2023} is a convex ${(N^{2}-1)}$-dimensional region with axes corresponding to Hermitian-operator basis elements and a boundary shape governed by their commutation relations.
As noise affects a quantum state, it distorts and shrinks this region, altering measured values.
In \cite{vovrosh_simple_2021}, the authors note that the simplicity of global depolarising noise
allows the distortion to be inverted, correcting the mean at the expense of increasing the sample variance.
Decomposing ${\hat{O}}$ in a basis of measurement operators \cite{rubin_application_2018} could potentially provide similar expressions for a wider range of noise models.
For general probabilistic unitary noise, however, the expectation value
\begin{align}
    \langle \hat{O} \rangle_{\mathcal{E}\left(\hat{\rho}\right)}
    &=
    \mathrm{Tr}
    \left[
        \hat{O}
        \left(
            \int d\xi
            \:
            \hat{U}_{\xi}
            \:
            \hat{\rho}
            \:
            \hat{U}^{\dagger}_{\xi}
        \right)
    \right]
    \nonumber\\
    &=
    \mathrm{Tr}
    \left[
        \left(
            \int d\xi
            \:
            \hat{U}^{\dagger}_{\xi}
            \:
            \hat{O}
            \:
            \hat{U}_{\xi}
        \right)
        \hat{\rho}
    \right]
    \label{eq:noisy_operator_expectation_trace}
\end{align}
becomes an arbitrary mixture over the basis elements, and can therefore not be reconstructed without knowledge of the full state ${\hat{\rho}}$.

The Stratonovich--Weyl conditions (Appendix~\ref{sec:the_stratonovich_weyl_conditions}) require that the expectation value of a basis operator can be expressed as
\begin{align}
    \langle \hat{O} \rangle
    &=
    \int d\xi \:
    W_{\hat{O}} (\xi)
    W_{\hat{\rho}} (\xi)
    .
\end{align}
In the spin Wigner function picture Equation~\eqref{eq:noisy_operator_expectation_trace} becomes
\begin{align}
    \langle \hat{O} \rangle_{\mathcal{E}\left(\hat{\rho}\right)}
    &=
    \int d\xi \:
    W_{\hat{O}} (\xi)
    \left(
        \int dy
        f (y\cdot \xi_{0})
        W_{\hat{\rho}} (y^{-1} \cdot \xi)
    \right)
    \nonumber\\
    &=
    \int d\xi \:
    \left(
        \int dy
        f (y\cdot \xi_{0})
        W_{\hat{O}} (y\cdot \xi)
    \right)
    W_{\hat{\rho}} (\xi)
    ,
\end{align}
with ${f (y\cdot \xi_{0})}$ a probability density function for erroneous rotations as described in Section~\ref{sec:spin_wigner_functions_and_spherical_harmonics}.

From the form for the kernel in Equation~\ref{eq:algebra_kernel_general} we see that,
just as expected from the operator representation, the number of additional basis operators
that must be measured to rescale the expectation value corresponds to the size of the equivalence
class of such operators under the action of the noise operator algebra.
While we could therefore work directly in terms of these operator bases, the spin Wigner function
picture becomes convenient when the noise may be expressed concisely in terms of harmonics ${Y_{n,j}}$
rather than in terms of basis operators.
In particular, as we note in Appendix~\ref{sec:the_degree_wise_convolution_of_spherical_harmonics},
when a noise distribution ${f (y\cdot \xi_{0})}$ is symmetric with respect to rotations about some
principal axis, its convolution may be applied element-wise.
Such symmetry indicates that the noise has a depolarising structure over some subspace
(all rotations in ${\mathrm{SU}(N-1)}$ about this axis have equal probability),
and as a result the expectation values for basis operators with support in this subspace remain
independent.

The simplest examples are cases of pure depolarising channels.
In the global depolarising model, with probability ${1-p}$ the state is left unchanged, while with probability ${p}$ the state is reduced to the maximally mixed state. The maximally mixed state consists only of the zeroth-order harmonic, so that the effect of depolarising noise is to re-scale all other harmonic coefficients by ${\left(1-p\right)}$.
For this model the parameterisation of the Wigner function is irrelevant.
For local depolarising noise the coefficients associated with all non-trivial basis operators or harmonics for qudit ${k}$ are reduced by a factor ${p_{k}}$.
This leads to exponential decay in expectation values according to the number of qudits in the support of each basis operator.
Labelling the set of qubits in the support of operator ${\hat{O}_{i}}$ by ${S_{\hat{O}_{i}}}$, the scale factor for the coefficient of ${\hat{O}_{i}}$ is ${\prod_{k\in S_{\hat{O}_{i}}}\left(1-p_{k}\right)}$.

Beyond depolarising noise, the next level of complexity arises from noise depending on a single angular variable of zero mean in each of several tensor-product factor spaces.
This is the case for the first two examples of Section~\ref{sub:noise_in_a_lie_subgroup}.
Since the standard convolution theorem applies to functions on the circle, we have that the corresponding circular harmonics are reduced element-wise, and the operator coefficients therefore decay independently of one another.
Re-scaling operator expectation values becomes a matter of determining this rate of decay from the coefficients ${f^{(i)}_{(n,j)}}$ of each of the basis elements in our observable of interest.

As a brief example of a tractable circular noise process, consider again the exchange interaction discussed in
Section~\ref{subs:charge_preserving_transformations}.
If we remove the dephasing terms and assume that the rate of exchange is state independent, then within each constant-excitation subspace the small-angle exchange operator is represented by a tridiagonal Toeplitz matrix with eigenvalues \cite{noschese_tridiagonal_2013}
\begin{align}
    \lambda_{h} (\theta / n \ll 1)
    =
    1 +
    i 2\frac{\theta}{n}
    \cos \left( \frac{h\pi}{D_{J}+1} \right)
    &
    ,
    \nonumber\\
    1\leq h \leq D_{J} &
    .
\end{align}
For larger angles this implies
\begin{align}
    \lambda_{h}(\theta)
    &=
    \lim_{n\rightarrow\infty}
    \lambda^{n}_{h} (\theta / n)
    =
    e^{
        i 2 \theta
        \cos \left( \frac{h\pi}{D_{J}+1} \right)
    }
    .
\end{align}
Now, just as in Section~\ref{sec:single_qubit_dephasing} we construct the states
\begin{align}
    \lvert \theta \rangle
    &=
    \bigoplus_{\substack{J\\1\leq h\leq D_{J}}}
    \left[
        \lambda_{h}(\theta)
    \right]
    ,
\end{align}
in the eigenbasis of the exchange operator,
and use the circular harmonics
${\sin(n\theta)}$, ${\cos(n\theta)}$
to define pseudo-displacement operators
\begin{align}
    \hat{D}_{n,f\in\{\sin,\cos\}}
    &=
    \int d\theta \:
    f(n\theta)
    \lvert \theta \rangle
    \langle \theta \rvert
    .
\end{align}
From these we derive coefficients for sets of traceless Hermitian basis
elements related to one another through the action of the exchange operator.
With the kernel thus defined, and supposing the noise distribution
is Gaussian for simplicity, the Fourier transform
would return a predictable Gaussian shape in the frequency basis,
and this could be used to re-weight the circular-harmonic coefficients of ${W_{\hat{O}} (\theta)}$.

\section{Discussion}
\label{sec:discussion}

In this work we have explored spin Wigner function parameterisation to efficiently represent and/or decompose the effects of noise on a quantum state or operator. Our results are built around the key observation that, since the noise operator algebra adopts the block diagonal structure represented in Equation~\ref{eq:tensor_sum_algebra_decomposition}, the spin coherent state construction pioneered by Brif and Mann \cite{brif_phase-space_1999} may be separated for each independent subspace.
Using spin coherent states defined over smaller subspaces allows us to maintain the Stratonovich--Weyl conditions while reducing the number of parameters required to represent probabilistic unitary noise.
In this manner we unify the description of several proposed kernels in the literature under a general form in Equation~\ref{eq:general_kernel_form}, which allows the substitution of any traceless Hermitian operator basis.
We describe explicit efficient kernel constructions for several noise processes that are local or slightly-entangling, or which satisfy specific rotational symmetries.

We expect the additional freedom we have introduced in the choice of parameterisation to find application in the
error mitigation problem of
inverting noise maps to correct observable expectation values \cite{vovrosh_simple_2021}.
Near term quantum systems often display noise that is either highly biased
\cite{tuckett_tailoring_2019}
or that quickly approaches a depolarising channel as circuit depth increases
\cite{tsubouchi_universal_2023}.
Though we find reflections of several important properties that also appear in the operator-basis picture,
we conjecture that spin Wigner functions could be useful in identifying depolarised subspaces in inhomogeneous circuits, for which the inverse map simplifies.

We identify circular and rotationally symmetric noise distributions as particularly convenient for the rescaling of harmonic coefficients.
A typical application of zero-noise extrapolation \cite{majumdar_best_2023} approximates circuit noise as a single-parameter function, and we believe it could be an interesting question to explore the relationship between the accuracy of zero-noise extrapolation and symmetries in the noise distribution of the kind discussed in this work.

There are a number of directions that could be pursued as extensions of this initial study.
For instance, though we have focused on the generalised Gell-Mann matrices as an operator basis,
different bases may be more or less convenient for the description of different noise processes.
We also leave to future work the question of other conditions under which the element-wise convolution theorem might be applied, or for which the more general degree-wise convolution discussed in Appendix~\ref{sec:the_degree_wise_convolution_of_spherical_harmonics} might be further restricted.
Finally, kernel constructions exploiting a tensor product structure are capable of expressing correlated rotations between subsystems, and so another area of interest will be the construction of logical spin Wigner functions for encoded quantum states, perhaps built on a modular subsystem decomposition of the kind introduced in \cite{Pantaleoni_2020}.

\begin{acknowledgments}
    We acknowledge financial support from the Samsung GRC project, the UK Hub in Quantum Computing and Simulation
    with funding from UKRI EPSRC grant EP/T001062/1, and EPSRC Distributed Quantum Computing and Applications grant EP/W032643/1.
\end{acknowledgments}

\bibliographystyle{apsrev4-1}
\bibliography{bibliography}

\appendix

\section{The Stratonovich--Weyl Conditions}
\label{sec:the_stratonovich_weyl_conditions}

The spin-Wigner function is a complete representation of an $N$-dimensional quantum state defined as a kernel transformation of the density matrix to satisfy the following key properties \cite{brif_phase-space_1999,tilma_wigner_2016}:
\begin{description}
    \item[S-W.1]
    ${W_{\hat{\rho}} (\xi) = \text{Tr} [ \hat{\rho}\hat{\Delta}(\xi)]}$
    and
    ${\hat{\rho} = \int W_{\hat{\rho}} (\xi)\hat{\Delta}(\xi) \: d\xi}$.
    
    \item[S-W.2]
    ${W_{\hat{\rho}}(\xi) \in \mathbb{R}}$
    (i.e. ${\hat{\Delta}(\xi)}$ is Hermitian).
    
    \item[S-W.3]
    ${\int W_{\hat{\rho}}(\xi) \: d\xi = \text{Tr} [\hat{\rho}]}$
    and
    ${\int \hat{\Delta}(\xi) d\xi = \mathds{1}}$. 
    
    \item[S-W.4]
    ${\int_\xi W_{\hat{\rho}}(\xi) W_{\hat{\rho}^{\prime}}(\xi) \: d\xi = \text{Tr} [ \hat{\rho} \hat{\rho}^{\prime} ]}$.
    
    \item[S-W.5]
    Any rotation of coordinates ${\xi'=R\xi}$ corresponds to conjugation of the transform kernel by a unitary operation:
    ${\hat{\Delta}\left( R\xi \right)=\hat{U}_{R} \hat{\Delta}\left( \xi \right) \hat{U}^{\dagger}_{R}}$.
    Therefore,
    ${W_{\hat{\rho}} \left( R\xi \right)= W_{\hat{U}^{\dagger}_{R}\hat{\rho}\hat{U}_{R}} \left( \xi \right)}$.
\end{description}
The implications of these requirements for the kernel ${\hat{\Delta}(\xi)}$ can be explicitly stated as \cite{brif_phase-space_1999}
\begin{align}
    \hat{\Delta}(\xi)
    &=
    \hat{\Delta}^{\dagger}(\xi)
    \nonumber\\
    \int
    d\xi\:
    \hat{\Delta}(\xi)
    &=
    \mathds{1}
    \nonumber\\
    \hat{\Delta}(R \xi)
    &=
    \hat{U}_{R}
    \hat{\Delta}(\xi)
    \hat{U}^{\dagger}_{R}
    ,
\end{align}
with ${R}$ a coordinate rotation and ${\hat{U}_{R}}$ a unitary operator.

\section{Convolution of Spherical Functions}
\label{sec:the_degree_wise_convolution_of_spherical_harmonics}

\subsection{Convolution is Applied Degree-wise}

Suppose we have the convolution of two functions $f$ and $g$ on ${S^{p-1}}$,
\begin{align}
    \left(
        f \ast g
    \right)
    \left( x \right)
    &=
    \frac{1}{\Omega_{\mathcal{R}_{p-1}}}
    \int_{\mathcal{R}_{p}}
    {
        f \left( y \right)
        g \left( y^{-1} x \right)
        \: dy
    }
    ,
\end{align}
where these functions have the respective harmonic expansions:
\begin{align}
    f
    &=
    \sum^{n^{\left(f\right)}_{max}}_{n=0}
    {
        \sum^{N\left( p , n \right)}_{j=1}
        {
            c^{\left(f\right)}_{n,j} Y_{n,j}
        }
    },
    \quad
    g
    =
    \sum^{n^{\left(g\right)}_{max}}_{m=0}
    {
        \sum^{N\left( p , m \right)}_{k=1}
        {
            c^{\left(g\right)}_{m,k} Y_{m,k}
        }
    }
    .
\end{align}

It can be shown that
inversion of the argument ${y^{-1} x}$ corresponds to a fixed orthogonal transformation, and so
to a degree-preserving linear transformation of the harmonic basis. We may therefore write the convolution as
\begin{align}
    \frac{1}{\Omega_{\mathcal{R}_{p-1}}}
    \sum_{n,m,j,k,l}
        c^{\left(f\right)}_{n,j} 
        c^{\left(g\right)}_{m,k}
        C_{l,k}
        \int_{\mathcal{R}_{p}}
        {
            Y_{n,j} \left( y \right)
            Y_{m,l} \left( x^{-1} y \right)
            \: dy
        }
        \label{eq:initial_inverted_convolution}
\end{align}
for some ${C_{l,k}}$, where the summation bounds are given by
\begin{align}
    \sum_{n,m,j,k,l}
    \sim 
    \sum^{n^{\left(f\right)}_{max}}_{n=0}
    \sum^{n^{\left(g\right)}_{max}}_{m=0}
    \sum^{N\left( p , n \right)}_{j=1}
    \sum^{N\left( p , m \right)}_{k=1}
    \sum^{N\left( p , m \right)}_{l=1}.
\end{align}
For any coordinate rotation ${R}$ we have \cite{efthimiou_spherical_2014}
\begin{align}
    Y_{n,j}\left(R\xi\right)
    &=
    \sum^{N\left(p,n\right)}_{l=0}
    {
        A_{l,j}
        Y_{n,l}\left(\xi\right)
    }
    ,
\end{align}
where ${\left[A_{l,j}\right]}$ is itself an orthogonal matrix.
Removing the ${x^{-1}}$ rotation from the argument in this way gives us
\begin{align}
    \frac{1}{\Omega_{\mathcal{R}_{p-1}}}
    \sum_{n,m,j,k,l}
        c^{\left(f\right)}_{n,j} 
        c^{\left(g\right)}_{m,k}
        A^{\left(x\right)}_{l,k}
        \int_{\mathcal{R}_{p}}
        {
            Y_{n,j} \left( y \right)
            Y_{m,l} \left( y \right)
            \: dy
        }
\end{align}
for some orthogonal ${[A^{\left(x\right)}_{l,k}]}$. In other words, defining
\begin{align}
    f_{n}
    &\defeq
    \sum^{N\left( p , n \right)}_{j=1}
    c^{\left(f\right)}_{n,j}
    Y_{n,j}
    \quad \text{and} \quad
    g_{n}
    \defeq
    \sum^{N\left( p , n \right)}_{j=1}
    c^{\left(g\right)}_{n,j}
    Y_{n,j}
    ,
\end{align}
we have that
\begin{align}
    \left(
        f \ast g
    \right)
    \left( x \right)
    &=
    \sum^{n_{max}}_{n=0}
    {
        \left(
            f_{n} \ast g_{n}
        \right)
        \left( x \right)
    }
    .
    \label{eq:generalised_convolution_theorem}
\end{align}

\subsection{The Convolution Theorem Applies for Certain Rotationally-Symmetric Functions}

In many cases we will be interested in noise distributions that are invariant under rotations
about the reference point $\xi_{0}$
(i.e. rotations in the hyper-plane normal to the reference point vector).
The (real-valued) function then depends only on
${\left\langle y^{-1} x \xi_{0} , \xi_{0} \right\rangle}$,
and we will be able to directly substitute
${y^{-1}x\rightarrow x^{-1}y}$.
Further, we can decompose each degree-$n$ harmonic according to the following theorem:
\begin{description}
    \item[Theorem (from \cite{efthimiou_spherical_2014})]
    For any spherical harmonic
    ${Y_{n}\left(\xi\right)}$
    of degree $n$, there exist coefficients $a_{k}$ and unit vectors $\eta_{k}$ such that
    \begin{align}
        Y_{n}\left(\xi\right)
        &=
        \sum^{N\left(p,n\right)}_{k=1}
        {
            a_{k}
            P_{n}\left( \left\langle \xi , \eta_{k} \right\rangle \right)
        }
        ,
        \label{eq:expand_harmonics_as_legendre_sum}
    \end{align}
    where
    ${P_{n}\left( \left\langle \xi , \eta \right\rangle \right)}$
    is the generalised Legendre polynomial of degree $n$ in $p$ dimensions (sometimes called a Gegenbauer or ultraspherical polynomial).
\end{description}
Taking
${\eta_{k}\rightarrow \xi_{0}}$,
we find that each integral in the convolution is equivalent to
\begin{align}
    \int_{\mathcal{R}_{p}}
    {
        Y_{n,j} \left( y \right)
        P_{n}\left( \left\langle y \xi_{0} , x \xi_{0} \right\rangle \right)
        \: dy
    }
    ,
\end{align}
which yields a simple multiple of
${Y_{n,j} \left( x \right)}$,
because for any spherical harmonic
${Y_{n}\left(\xi\right)}$ \cite{efthimiou_spherical_2014},
\begin{align}
    Y_{n}\left(\xi\right)
    &=
    \frac{N\left(p,n\right)}{\Omega_{p-1}}
    \int_{S^{p-1}}
    {
        Y_{n}\left(\eta \right)
        \: P_{n}\left( \left\langle \xi , \eta \right\rangle \right)
        \: d\eta
    }
    \nonumber\\
    \Omega_{p-1}
    &=
    \int_{S^{p-1}}
    {
        1
        \: d\eta
    }
    .
\end{align}

The standard convolution theorem therefore applies.
For functions of this type, the choice of basis ${Y_{n,j}}$ is irrelevant; we always have ${c_{n,j}=a_{n}\frac{\Omega_{p-1}}{N\left(p,n\right)}, \forall j}$.
On $S^{2}$, these distributions correspond to harmonic expansions containing only `zonal' harmonic functions.
On $S^{1}$, all functions obey the standard convolution theorem.

\section{The Harmonic Displacement Kernel Construction}
\label{sec:the_harmonic_displacement_kernel_construction}

In \cite{brif_phase-space_1999}, a quite general construction for the spin Wigner function kernel ${\hat{\Delta}(\xi)}$ was proposed in terms of spin coherent states and spherical harmonics. In this paper we use an alternative convention for the spherical harmonics that leaves them real rather than complex, so we briefly recap the construction below under this convention. This leads to a few slight differences, such as the displacement-equivalent operators ${\hat{D}_{n,j}}$ becoming Hermitian.

Let ${G}$ be the group ${\mathrm{SU}(N)}$ of operations on a quantum state, and ${H}$ the isometry group ${\mathrm{SU}(N-1)}$ leaving a single representative state vector ${\left\lvert \psi_{0} \right\rangle}$ (which we will take to be the highest-weight state) invariant. The coset space ${X=G/H}$ defines the coordinate space for the spin coherent states ${\left\lvert \xi \right\rangle}$ \cite{nemoto_generalized_2000}, and is isomorphic to the ${(2N-1)}$-sphere:
$${
	\frac{
		\mathrm{SU}\left(N\right)
	}{
		\mathrm{SU}\left(N-1\right)
	}
	\cong
	S^{2N-1}
	.
}$$
The spherical harmonics ${Y_{n,j}(\xi)}$ have been discussed in Section~\ref{sec:spin_wigner_functions_and_spherical_harmonics}. Define the following operators:
$${
	\hat{D}_{n,j}
	=
	\int_{X} d\xi
	\:
	Y_{n,j}(\xi)
	\:
	\left\lvert \xi \right\rangle
	\left\langle \xi \right\rvert
	,
}$$
which are simply the coherent states weighted by a spherical harmonic. The claim is that the kernel
$${
	\hat{\Delta}(\xi)
	=
	C_{\Delta}
	\sum_{n,j}
	Y_{n,j}(\xi)
	\hat{D}_{n,j}
	,
}$$
with some constant normalising factor ${C_{\Delta}}$, satisfies the Stratonovich-Weyl conditions outlined in Appendix~\ref{sec:the_stratonovich_weyl_conditions}:
As the harmonics are real, it is straightforward to see that the kernel is Hermitian (as indeed are the ${\hat{D}_{n,j}}$ operators).
Integrating over the coordinates, normality is satisfied as
\begin{align}
	\int_{X} d\xi
	\:
	\hat{\Delta}(\xi)
	&=
	C_{\Delta}
	\sum_{n,j}
	\left(
		\int_{X} d\xi
		\:
		Y_{n,j}(\xi)
	\right)
	\hat{D}_{n,j}
    \nonumber\\
	&=
	C_{\Delta}
	Y^{-1}_{0,0}
	\hat{D}_{0,0}
    \nonumber\\
	&=
	C_{\Delta}
	\int_{X} d\xi
	\:
	\left\lvert \xi \right\rangle
	\left\langle \xi \right\rvert
    \nonumber\\
	&=
	\hat{\mathbb{I}}
	,
\end{align}
for some constant ${C_{\Delta}}$ (via the resolution of unity for the spin coherent states).
For the final condition of the kernel, covariance, we have
$${
	\hat{\Delta}(g \cdot \xi)
	=
	C_{\Delta}
	\sum_{n,j}
	Y_{n,j}(g \cdot \xi)
	\int_{X} d\eta
	\:
	Y_{n,j}(\eta)
	\:
	\left\lvert \eta \right\rangle
	\left\langle \eta \right\rvert
	.
}$$
Now, for any rotation ${g}$ of coordinates on the spherical coset space we have that \cite{efthimiou_spherical_2014}
$${
	Y_{n,j}(g\cdot \xi)
	=
	\sum_{l}
	C_{lj}
	Y_{n,l}(\xi)
}$$
for some orthogonal matrix ${[C_{lj}]}$ depending on ${g}$. Further expanding the kernel, this give us
\begin{align}
	\hat{\Delta}(g \cdot \xi)
	&=
	C_{\Delta}
	\sum_{n,j}
	\left(
		\sum_{l}
		C_{lj}
		Y_{n,l}(\xi)
	\right)
	\int_{X} d\eta
	\:
	Y_{n,j}(\eta)
	\:
	\left\lvert \eta \right\rangle
	\left\langle \eta \right\rvert
    \nonumber\\
	&=
	C_{\Delta}
	\sum_{n,l}
	Y_{n,l}(\xi)
	\int_{X} d\eta
	\:
	\left(
		\sum_{j}
		C_{lj}
		Y_{n,j}(\eta)
	\right)
	\:
	\left\lvert \eta \right\rangle
	\left\langle \eta \right\rvert
	.
\end{align}
The summation now appearing inside the integral corresponds to the transposed (i.e. inverse) transformation matrix. Noting that the integral is over the Haar measure, we obtain
\begin{align}
	\hat{\Delta}(g \cdot \xi)
	&=
	C_{\Delta}
	\sum_{n,l}
	Y_{n,l}(\xi)
	\int_{X} d\eta
	\:
	Y_{n,l}(g^{-1}\cdot \eta)
	\:
	\left\lvert \eta \right\rangle
	\left\langle \eta \right\rvert
    \nonumber\\
	&=
	C_{\Delta}
	\sum_{n,l}
	Y_{n,l}(\xi)
	\int_{X} d\eta
	\:
	Y_{n,j}(\eta)
	\:
	\left\lvert g \cdot \eta \right\rangle
	\left\langle g \cdot \eta \right\rvert
	,
\end{align}
or
\begin{align}
	\hat{\Delta}(g \cdot \xi)
	&=
	C_{\Delta}
	\sum_{n,l}
	Y_{n,l}(\xi)
	\left(
		\hat{U}_{g}
		\hat{D}_{n,j}
		\hat{U}^{\dagger}_{g}
	\right)
    \nonumber\\
	&=
	\hat{U}_{g}
	\hat{\Delta}(\xi)
	\hat{U}^{\dagger}_{g}
	,
\end{align}
so that the covariance requirement is satisfied.

\section{A Noise-Restricted Kernel}
\label{sec:app_07_noise_restricted_kernel}

Consider a single term from Equation~\eqref{eq:general_kernel_form},
\begin{align}
    \hat{O}_{i}
	\sum_{n,j}
	f^{(i)}_{(n,j)}
	Y_{n,j} \left( \xi \right)
    .
    \label{eq:kernel_single_term}
\end{align}
Our goal will be to decompose this term to extract the dependence on
a subset ${\omega}$ of the angles in ${\xi}$
(and we will denote the residual set of angles ${\eta=\xi / \omega}$).
We first note that the spherical harmonics ${Y_{n,j}}$ are homogeneous polynomials of degree ${n}$,
and we may therefore write
\begin{align}
    Y_{n,j} \left( \xi \right)
    &=
    \sum^{n}_{k=0}
    \sum^{L}_{l=1}
    g_{n-k,l}(\eta)
    h_{k,l}(\omega)
    \nonumber\\
    L
    &=
    \left(\begin{array}{c}
        k + \lvert \omega \rvert - 1\\
        k
    \end{array}\right)
    ,
\end{align}
where ${\lvert \omega \rvert}$ is the dimension of the space spanned by the angles ${\omega}$,
${g_{n-k}(\eta)}$ and ${h_{k}(\omega)}$ are real homogeneous polynomials of respective degrees ${n-k}$ and ${k}$,
and the range of ${l}$ is determined by the number of distinct monomial terms of degree ${k}$ over ${\lvert \omega \rvert}$
elements (the stars-and-bars expression).

Since ${Y_{n,j} \left( \xi \right)}$ is real and square-integrable and the ${h_{k,l}(\omega)}$ are linearly independent monomial terms of maximum degree ${n}$, ${h_{k,l}(\omega)}$ are also real and square integrable.
They may be therefore be expressed again as linear combinations of spherical harmonics of degree ${k}$:
\begin{align}
    h_{k,l}(\omega)
    &=
    \sum_{m}
    c^{(l)}_{k,m}
    Y_{k,m}(\omega)
    .
\end{align}
Looking back at the initial Equation~\eqref{eq:kernel_single_term}, we now have
\begin{align}
	\hat{O}_{i}
	\sum_{n,j}
	f^{(i)}_{(n,j)}
	\sum^{n}_{k=0}
    \sum^{L}_{l=1}
    g_{n-k,l}(\eta)
    \sum_{m}
    c^{(l)}_{k,m}
    Y_{k,m}(\omega)
	,
\end{align}
and we may re-arrange the order of summation to obtain
\begin{align}
	\hat{O}_{i}
    \sum_{k,m}
    \left(
    	\sum_{n\geq k,j}
        \sum^{L}_{l=1}
        c^{(l)}_{k,m}
    	f^{(i)}_{(n,j)}
        g_{n-k,l}(\eta)
    \right)
    Y_{k,m}(\omega)
	.
\end{align}

The ${g_{n-k,l}(\eta)}$ may also be decomposed in spherical harmonics,
and making the replacement ${n-k \rightarrow n'}$ gives
\begin{align}
	\hat{O}_{i}
    \sum_{k,m}
    &
    \left(
    	\sum_{n',p}
        q^{(i)}_{(k,m,n',p)}
        Y_{n',p}(\eta)
    \right)
    Y_{k,m}(\omega)
    \nonumber\\
    q^{(i)}_{(k,m,n',p)}
    &=
    \sum^{L}_{l=1}
    c^{(l)}_{k,m}
    c^{(l)}_{n',p}
    \sum_{j}
    f^{(i)}_{(n'+k,j)}
	.
\end{align}
This is the form of the decomposition used in
Equation~\eqref{eq:algebra_kernel_general}.

We know that the coefficients of distinct operators are orthogonal on integration
over all angular parameters. This implies that
\begin{align}
    r^{(i)}_{(k,m)}
    &=
    \sum_{n',p}
    q^{(i)2}_{(k,m,n',p)}
\end{align}
are discrete orthogonal functions when summed over ${k}$ and ${m}$.

\section{Orthogonality Relations for a Tensor-Sum of Spin Coherent States}
\label{sec:orth}

In Section~\ref{subs:charge_preserving_transformations} we introduced pseudo-displacement operators based on
the tensor sum of ${\mathrm{SU}(2)}$ spin coherent states of varying dimension, noting that coefficient expansions in these
operators could only be taken within the parenthetical sums of Equation~\eqref{eq:algebra_kernel_general},
rather than over the full set of basis operators.
In this appendix, we confirm that for the pseudo-displacement operators constructed from
Equation~\eqref{eq:tensor_sum_su2_spin_coh_states},
the orthogonality relations \cite{tilma_su_2012}
\begin{align}
    \int d\xi \:
    \left\langle \xi \right\rvert
        \hat{\Lambda}_{i}
    \left\lvert \xi \right\rangle
    \left\langle \xi \right\rvert
        \hat{\Lambda}_{k}
    \left\lvert \xi \right\rangle
    &\propto
    \delta_{ik}
    ,
\end{align}
now only hold within restricted subsets ${\{\hat{\Lambda}_{k}\}}$.

\subsection{General Orthogonality Relations}

For
${G}$
    a compact group,
${\pi^{\alpha}(g\in G)}$
    a complete set of irreducible representations
    with dimensions ${d_{\alpha}}$
and
${\phi^{\alpha}_{v,w}(g) = \langle v , \pi^{\alpha}(g) w \rangle }$
    the matrix coefficients for representation ${\alpha}$,
the Schur orthogonality relations tell us
\begin{enumerate}
    \item
    If ${\pi^{\alpha} \not\cong \pi^{\beta}}$,
    \begin{align}
        \int_{G} dg \:
            \phi^{\alpha}_{v,w}(g)
            \phi^{\beta}_{v',w'}(g)
        &=
        0
        .
    \end{align}
    \item
    If ${\{e_{i}\}}$ is an orthonormal basis for ${\pi^{\alpha}}$,
    \begin{align}
        \int_{G} dg \:
            \phi^{\alpha}_{e_{i},e_{j}}(g)
            \phi^{\alpha *}_{e_{k},e_{l}} (g)
        &=
        \delta_{i,k}
        \delta_{j,l}
        d^{-1}_{\alpha}
        .
    \end{align}
\end{enumerate}

In particular, for a basis of states ${\lvert j \rangle}$ in a Hilbert space of dimension ${N}$ and
(irreducible) matrix representations ${U_{\xi}}$ for elements ${\xi}$ of ${\mathrm{SU}(K\leq N)}$,
the Schur orthogonality relations indicate
\begin{align}
    \int_{\xi} d\xi \:
        \langle N \rvert
            U^{\dagger}_{\xi}
        \lvert j \rangle
        \langle k \rvert
            U_{\xi}
        \lvert N \rangle
    &=
    \int_{\xi} d\xi \:
        \langle \xi \vert j \rangle \langle k \vert \xi \rangle
    \nonumber\\
    &=
    \frac{
        \delta_{j,k}
    }{N}
    ,
\end{align}
as expected from the resolution of unity in the spin coherent states.
Next consider matrices ${M}$ in the Hilbert space of dimension ${N^{2}-1}$ spanned by the generalised Gell-Mann matrices
${\Lambda_{i}}$, with inner product ${\frac{1}{2}\mathrm{Tr}\left[\Lambda_{i} \Lambda_{j} \right]=\delta_{i,j}}$,
and with action ${M \rightarrow U_{\xi} M U^{\dagger}_{\xi} }$.
This larger space is a reducible representation of ${\mathrm{SU}(K)}$,
though the ${N}$-dimensional states were not.
We can however vectorize ${M}$ in a tensor product as
\begin{align}
    M
    &=
    \sum_{m,n}
    c_{m,n}
    \lvert m \rangle
    \langle n \rvert
    \nonumber\\
    &\rightarrow
    \sum_{m,n}
    c_{m,n}
    \lvert m \rangle
    \otimes
    \lvert n \rangle
    ,
\end{align}
where now the action of ${U_{\xi}}$ is given by
\begin{align}
    U_{\xi} M U^{\dagger}_{\xi}
    &=
    \sum_{m,n}
    c_{m,n}
    U_{\xi} \lvert m \rangle
    \langle n \rvert U^{\dagger}_{\xi}
    \nonumber\\
    &\rightarrow
    \sum_{m,n}
    c_{m,n}
    \left( U_{\xi} \lvert m \rangle \right)
    \otimes
    \left( U_{\xi} \lvert n \rangle \right)
    .
\end{align}
For ${\mathrm{SU}(2)}$, the decomposition of this tensor product structure into
irreducible subspaces is known and is given by
Wigner D-matrices of linearly increasing dimension, with elements weighted by the Clebsch--Gordon coefficients
\cite{sakurai_modern_1985}:
\begin{align}
    U^{(j)}_{\xi}\otimes U^{(j)}_{\xi}
    &=
    \bigoplus^{2j}_{k=0}
    U^{(k)}_{\xi}
    ,
    \nonumber\\
    \langle m_{1},m_{2} \rvert
        U^{(j)}_{\xi} \otimes U^{(j)}_{\xi}
    \lvert m^{\prime}_{1} , m^{\prime}_{2} \rangle
    &=
    \nonumber\\
    \sum^{2j}_{k=0}
    \sum_{m,m^{\prime}}
    C^{j,m_{1},m_{2},k,m}_{j,m^{\prime}_{1},m^{\prime}_{2},k,m^{\prime}}
    &
    \langle m \rvert U^{(k)}_{\xi} \lvert m^{\prime} \rangle
    ,
    \nonumber\\
    C^{j,m_{1},m_{2},k,m}_{j,m^{\prime}_{1},m^{\prime}_{2},k,m^{\prime}}
    &=
    \langle j ; m_{1},m_{2} \vert j ; k,m \rangle
    \nonumber\\
    &\qquad\cdot
    \langle j ; m^{\prime}_{1},m^{\prime}_{2} \vert j ; k,m^{\prime} \rangle
    .
\end{align}
On performing this decomposition, and for basis operators ${O_{j}}$ within a single irreducible subspace,
the Schur orthogonality relations
with
\begin{align}
    e_{i}
    &=
    e_{k}
    = \lvert N \rangle \langle N \rvert
    ,\quad
    e_{j} \rightarrow O_{j}
    ,\quad
    e_{l} \rightarrow O_{l}
\end{align}
indicate
\begin{align}
    &
    \int_{\xi} d\xi \:
        \mathrm{Tr}\left[ \langle N \rvert U^{\dagger}_{\xi} O_{j} U_{\xi} \lvert N \rangle \right]
        \mathrm{Tr}\left[ \langle N \rvert U^{\dagger}_{\xi} O_{l} U_{\xi} \lvert N \rangle \right]
    \nonumber\\
    &=
    \int_{\xi} d\xi \:
        \langle \xi \rvert O_{j} \lvert \xi \rangle
        \langle \xi \rvert O_{l} \lvert \xi \rangle
    \nonumber\\
    &\propto
    \delta_{j,l}
    .
\end{align}
However, operators in different irreducible subspaces, and transition operators between them,
are not guaranteed to be orthogonal under the action of ${U_{\xi}}$.
The range of integration is over the full group ${\mathrm{SU}(K)}$, including the
isotropy subgroup leaving the initial state ${\lvert N \rangle}$ invariant.

\subsection{The ${\mathrm{SU}(2)}$ Spin Coherent States}

First, denoting the ${m}$th basis element of the ${J}$-excitation subspace as ${\lvert J ; m \rangle}$,
our tensor sum of ${\mathrm{SU}(2)}$ spin coherent states with coordinates
${\xi = \lvert \phi_{1}, \phi_{2}, \theta \rangle}$
becomes \cite{nemoto_generalized_2000}
\begin{align}
    \lvert \xi \rangle
    &=
    \sum_{\substack{J\\0\leq m \leq D_{J}-1}}
    \eta^{D_{J}}_{m} ( \xi )
    \lvert J ; m \rangle
    ,
    \nonumber\\
    \eta^{D_{J}}_{m} ( \xi )
    &=
    e^{im\phi_{2}}
    e^{i(D_{J}-m-1)\phi_{1}}
    \sin^{m}(\theta)
    \nonumber\\
    &\qquad \cdot
    \cos^{D_{J}-m-1}(\theta)
    \begin{pmatrix}
        D_{J}-1 \\ m
    \end{pmatrix}^{1/2}.
\end{align}

Next, for convenience of notation, express the off-diagonal generalised Gell-Mann matrices as
\begin{align}
    X^{J,m}_{K,n}
    &=
    \lvert J ; m \rangle
    \langle K ; n \rvert
    +
    \lvert K ; n \rangle
    \langle J ; m \rvert
    \nonumber\\
    Y^{J,m}_{K,n}
    &=
    -i
    \lvert J ; m \rangle
    \langle K ; n \rvert
    +
    i
    \lvert K ; n \rangle
    \langle J ; m \rvert
    .
\end{align}
We have
\begin{align}
    \langle \xi \rvert
        X^{J,m}_{K,n}
    \lvert \xi \rangle
    &=
    \sum_{J',m'}
    \sum_{K'n'}
    (\eta^{D_{J'}}_{m'} ( \xi ))^{*}
    \eta^{D_{K'}}_{n'} ( \xi )
    \nonumber\\
    &\qquad\qquad\quad
    \cdot
    \left[
        \langle J' ; m' \vert J ; m \rangle
        \langle K ; n \vert K' ; n' \rangle
    \right.
    \nonumber\\
    &\qquad\qquad\qquad
    \left.
        +
        \langle J' ; m' \vert K ; n \rangle
        \langle J ; m \vert K' ; n' \rangle
    \right]
    \nonumber\\
    &=
    ( \eta^{D_{J}}_{m} ( \xi ) )^{*}
    \eta^{D_{K}}_{n} ( \xi )
    +
    \eta^{D_{J'}}_{m'} ( \xi )
    ( \eta^{D_{K'}}_{n'} ( \xi ) )^{*}
    \nonumber\\
    &=
    2 \: \mathrm{Re}
    \left[
        ( \eta^{D_{J}}_{m} ( \xi ) )^{*}
        \eta^{D_{K}}_{n} ( \xi )
    \right]
    ,
    \nonumber\\
    \langle \xi \rvert
        Y^{J,m}_{K,n}
    \lvert \xi \rangle
    &=
    \sum_{J',m'}
    \sum_{K'n'}
    (\eta^{D_{J'}}_{m'} ( \xi ))^{*}
    \eta^{D_{K'}}_{n'} ( \xi )
    \nonumber\\
    &\qquad\qquad\quad
    \cdot
    i
    \left[
        -
        \langle J' ; m' \vert J ; m \rangle
        \langle K ; n \vert K' ; n' \rangle
    \right.
    \nonumber\\
    &\qquad\qquad\qquad
    \left.
        +
        \langle J' ; m' \vert K ; n \rangle
        \langle J ; m \vert K' ; n' \rangle
    \right]
    \nonumber\\
    &=
    -i( \eta^{D_{J}}_{m} ( \xi ) )^{*}
    \eta^{D_{K}}_{n} ( \xi )
    +
    i
    \eta^{D_{J'}}_{m'} ( \xi )
    ( \eta^{D_{K'}}_{n'} ( \xi ) )^{*}
    \nonumber\\
    &=
    2 \: \mathrm{Im}
    \left[
        ( \eta^{D_{J}}_{m} ( \xi ) )^{*}
        \eta^{D_{K}}_{n} ( \xi )
    \right]
    .
\end{align}

The ranges of integration for the state parameters
${\phi_{1}, \phi_{2}, \theta}$
are
\begin{align}
    \phi_{1}, \phi_{2} &\in [ 0, 2\pi ] \\
    \theta &\in [ 0, \pi/2 ]
    .
\end{align}
Since only ${\phi_{1}}$ and ${\phi_{2}}$ are integrated over the full unit circle,
these variables govern orthogonality and so
we drop the constant factors and
${\theta}$-dependence in what follows.
We have
\begin{align}
    \mathrm{Re}
    \left[
        ( \eta^{D_{J}}_{m} ( \xi ) )^{*}
        \eta^{D_{K}}_{n} ( \xi )
    \right]
    &\propto
    \cos( \delta^{n}_{m} \phi_{2} )
    \cos( \Delta^{K,n}_{J,m} \phi_{1} )
    \nonumber\\
    &\qquad
    +
    \sin( \delta^{n}_{m} \phi_{2} )
    \sin( \Delta^{K,n}_{J,m} \phi_{1} )
    ,
    \nonumber\\
    \mathrm{Im}
    \left[
        ( \eta^{D_{J}}_{m} ( \xi ) )^{*}
        \eta^{D_{K}}_{n} ( \xi )
    \right]
    &\propto
    \cos( \delta^{n}_{m} \phi_{2} )
    \sin( \Delta^{K,n}_{J,m} \phi_{1} )
    \nonumber\\
    &\qquad
    +
    \sin( \delta^{n}_{m} \phi_{2} )
    \cos( \Delta^{K,n}_{J,m} \phi_{1} )
    ,
    \nonumber\\
    \delta^{n}_{m}
    &=
    n - m
    ,
    \nonumber\\
    \Delta^{K,n}_{J,m}
    &=
    D_{K} - D_{J} - \delta^{n}_{m}
    .
\end{align}

We first note that any ${X/Y}$ pair is orthogonal,
since ${\delta^{n}_{m},\Delta^{K,n}_{J,m} \in \mathbb{Z}}$ and so
\begin{align}
    0
    &=
    \int^{2\pi}_{0} d\phi_{2} \:
    \cos( \delta^{n}_{m} \phi_{2} )
    \sin( \delta^{n'}_{m'} \phi_{2} )
    \nonumber\\
    &=
    \int^{2\pi}_{0} d\phi_{1} \:
    \cos( \Delta^{K,n}_{J,m} \phi_{1} )
    \sin( \Delta^{K',n'}_{J',m'} \phi_{1} )
    .
\end{align}
This condition is necessary, because factors
${e^{i\phi_{1}\hat{J}^{(1)}_{Z}}}$
and
${e^{i\phi_{2}\hat{J}^{(2)}_{Z}}}$
in Equation~\eqref{eq:example_3_charge_preserving_transformations}
cause transitions between
${X^{J,m}_{K,n}}$ and ${Y^{J,m}_{K,n}}$.

Next, for operators
${X^{J,m}_{K,n}}$ and ${X^{J',m'}_{K',n'}}$
we have
\begin{align}
    &
    \mathrm{Re}
    \left[
        ( \eta^{D_{J}}_{m} ( \xi ) )^{*}
        \eta^{D_{K}}_{n} ( \xi )
    \right]
    \mathrm{Re}
    \left[
        ( \eta^{D_{J'}}_{m'} ( \xi ) )^{*}
        \eta^{D_{K'}}_{n'} ( \xi )
    \right]
    \nonumber\\
    &\propto
    \quad
    \cos( \delta^{n}_{m} \phi_{2} )
    \cos( \Delta^{K,n}_{J,m} \phi_{1} )
    \cos( \delta^{n'}_{m'} \phi_{2} )
    \cos( \Delta^{K',n'}_{J',m'} \phi_{1} )
    \nonumber\\
    &\qquad
    +
    \sin( \delta^{n}_{m} \phi_{2} )
    \sin( \Delta^{K,n}_{J,m} \phi_{1} )
    \sin( \delta^{n'}_{m'} \phi_{2} )
    \sin( \Delta^{K',n'}_{J',m'} \phi_{1} )
    ,
    \label{eq:orthogonality_appendix_double_x}
\end{align}
so that the integral is only non-zero when
\begin{align}
    \delta^{n}_{m}
    &=
    \pm
    \delta^{n'}_{m'}
    \quad \text{and} \quad
    \Delta^{K,n}_{J,m}
    =
    \pm
    \Delta^{K',n'}_{J',m'}
    .
\end{align}
In particular, for transition operators between two
constant-excitation subspaces ${J}$ and ${K}$ (which may be equal),
the orthogonality relations hold for all X-type lambda-matrices
along the ${k}$th anti-diagonal, row, \emph{or} column, for any ${k}$.

Not every combination ${K,J,n,m,K',J',n',m'}$ in
Equation~\eqref{eq:orthogonality_appendix_double_x}
results in an orthogonality relation.
This implies that the noise variables alone, and
their pseudo-displacement operator,
are not sufficient to construct a set of orthogonal coefficients
over the full set of basis operators in the larger ${N}$-dimensional space,
as expected from Equation~\eqref{eq:algebra_kernel_general}.

\end{document}